\documentstyle[12pt,aaspp4,psfig]{article}   

\lefthead{HARDING}
\righthead{PULSAR X-RAY AND GAMMA-RAY PROFILES}
\newcommand{\be}{\begin{equation}} 
\newcommand{\ee}{\end{equation}} 

\newcommand{\tnm}{\tablenotemark}
\newcommand{\tnt}{\tablenotetext}

\def\lsim{\lower 2pt \hbox{$\, \buildrel {\scriptstyle <}\over
         {\scriptstyle \sim}\,$}}
\newcommand\gsim{\buildrel > \over \sim}

\def\pmb#1{\setbox0=\hbox{#1}%
  \kern-0.015em\copy0\kern-\wd0
  \kern0.03em\copy0\kern-\wd0
  \kern-0.015em\raise0.0433em\box0 }

\begin{document}
\newcommand{\figureout}[3]{\psfig{figure=#1,width=7in,angle=#2} 
  \figcaption{#3} }  

\title{PULSAR X-RAY AND GAMMA-RAY PULSE PROFILES: 
CONSTRAINT ON OBLIQUITY AND OBSERVER ANGLES}
\author{Alice K. Harding and Alexander G. Muslimov
        \footnote{NRC/NAS Senior Research Associate}
}

\affil{Laboratory for High Energy Astrophysics \\
       NASA/Goddard Space Flight Center, Code 661 \\
       Greenbelt, MD 20771 USA \\
       harding@twinkie.gsfc.nasa.gov; muslimov@lhea1.gsfc.nasa.gov}

\begin{abstract}

We model the thermal X-ray profiles of Geminga, Vela and PSR 0656+14, 
which have also been detected as $\gamma $-ray pulsars, to constrain the 
phase space of obliquity and observer angles required to reproduce the 
observed X-ray pulsed fractions and pulse widths. These geometrical 
constraints derived from the X-ray light curves are explored for 
various assumptions about surface temperature distribution and
flux anisotropy caused by the magnetized atmosphere. We include 
curved spacetime effects on photon trajectories and magnetic field. 
The observed $\gamma $-ray pulse profiles
are double peaked with phase separations of 0.4 - 0.5 between the peaks.
Assuming that the $\gamma $-ray profiles are due to emission in a hollow 
cone centered on the magnetic pole, we derive the constraints on the 
phase space of obliquity and observer angles, for different $\gamma $-ray 
beam sizes, required to produce the observed $\gamma $-ray peak phase 
separations. We compare the constraints from the X-ray emission to
those derived from the observed $\gamma $-ray pulse profiles, and find
that the overlapping phase space requires both obliquity and observer 
angles to be smaller than $20 - 30^0$, implying $\gamma $-ray
beam opening angles of at most $30-35^0$.
               
\end{abstract}


\bigskip
\bigskip
Submitted to {\em The Astrophysical Journal.}

\newpage 
\section{INTRODUCTION
         \label{sec:1}}

The multifrequency observations of $\gamma $-ray pulsars may
potentially constrain the geometry and location of emission regions 
in a neutron star (NS) magnetosphere. Such a constraint is necessary 
for our understanding of the entire picture of pulsar emission in 
different energy bands. However, in practice, the interpretation of 
high-energy observations is rather ambiguous and involves a 
number of model assumptions. The recent X-ray observations of 
pulsars Geminga, Vela, and PSR 0656+14 seem to 
indicate that the pulsed emission has a two-component X-ray spectrum
(\cite{oge95}, \cite{hal97}, \cite{str97}), where the soft (pulsed) 
component is likely due to thermal emission.  Theoretical work by 
Pavlov et al. (1994) 
has shown that X-ray pulsation reaching the observed pulsed fractions
can result from anisotropic emission in a magnetized atmosphere, even
when the entire NS surface radiates.  If this
interpretation is correct, then the peak in the soft thermal X-ray 
profile is near the phase of closest approach to the magnetic pole.
It is thus interesting to explore the relation between X-ray
profiles as thermal 
black-body emission from the whole stellar surface ``beamed'' along 
the magnetic field in a NS atmosphere and non-thermal 
pulsed $\gamma $-ray emission from these
pulsars generated in the innermost magnetosphere of a NS (polar cap
model)(Harding \& Muslimov 1997a).  The polar cap models for pulsed
gamma-ray emission (Daugherty \& Harding 1996, Sturner \& Dermer 1994)
predict double-peaked pulse profiles, where the closest approach to
the pole is centered between the pulses.  The peak of the broad thermal
X-rays profile should thus also occur between the two gamma-ray pulses.

In this paper we model the soft X-ray and $\gamma$-ray light curves 
for the pulsars Geminga, Vela, and PSR 0656+14 surveying 
all possible orientations (see Figure 1) between the magnetic and 
spin axes (referred to as the {\bf obliquity} angles) and the angle 
between the observer's line of sight and spin axis (referred to
as the {\bf observer} angles). In our modeling of  the X-ray and $\gamma
$-ray light curves we assume that the NS has a centered dipole-like surface 
magnetic field, and that the thermal flux from the stellar surface  
corresponds to that expected in a cooling NS of age 
$\sim 10^4-3\times 10^5$ yr (the age category 
of Geminga, Vela and PSR 0656+14). In our analysis of the
main observational features we consciously avoid additional model 
assumptions such as the possibility of polar cap heating by the 
precipitating relativistic particles and $\gamma$-rays, 
off-setting of the magnetic axis of the NS and/or the presence of
higher-order multipoles on the stellar surface (which are not 
unreasonable at all for future modeling), and occurrence
of any specific cooling scenario (e.g. such as those with 
internal heating of the NS, or those allowing for any of 
the countless variants of rapid or slow cooling, etc).  Also, we 
have not considered very compact NS models for which the effects of 
strong gravity by themselves may result in quite interesting
signatures (\cite{shib95}). All these possibilities, being
attractive for a  
theoretical study, are rather abstract when discussed in the 
context of the currently available X-ray and $\gamma $-ray 
observational data on pulsars. The quality of the observations
and also the complexity of theoretical modeling (which usually 
involves many free parameters and poorly justified model assumptions) 
do not allow any conclusive statement regarding any of the 
abovementioned possibilities.

The main goal of our study is to demonstrate that a polar cap model 
for the interpretation of the $\gamma $-ray 
emission is  viable at least for Geminga, Vela, and PSR 0656+14,
and that the main observed X-ray and $\gamma $-ray pulse characteristics 
(X-ray pulsed fraction, half-width of the X-ray profile, phase 
separation of $\gamma $-ray peaks, and a phase shift $\sim 0.1 - 0.2$  
between the $\gamma $-ray peaks) for these
pulsars can be understood within the framework of a standard 
NS model with a dipole-like  magnetic field and a relatively small 
obliquity ($\lsim 30-35^o$).
We begin our paper (\S~2) with a summary of the observational 
X- and $\gamma $-ray data on pulsars Geminga, Vela, and PSR 0656+14. 
In $\S ~3$ we describe the details of our modeling of the 
observed soft X- and $\gamma $-ray emission, and in $\S ~4$ we 
present the results of our numerical calculations. Our principal 
conclusions are summarized in $\S ~5$.

\section{SUMMARY OF THE OBSERVATIONAL DATA
         \label{sec:2}         }

In Figure 2 we show soft X-ray and high-energy $\gamma$-ray pulse 
profiles for Geminga, Vela and PSR 0656+14. 
We have chosen these sources for modeling 
X-ray and $\gamma$-ray pulsed profiles because they have been identified as 
having well-defined thermal components.  Recent hard X-ray observations of
Geminga (Halpern \& Wang 1997) and PSR 0656+14 (Greiveldinger et al. 1996)
with ASCA and Vela with RXTE (Rossi X-ray Timing Explorer; 
Strickman, Harding \& De Jager 1998) have
clearly defined the existence of separate non-thermal components and 
therefore greatly strengthened the interpretation of the soft X-ray
components seen by ROSAT as thermal in origin.  ROSAT observations of these
sources had revealed phase shifts between the pulses seen in low energy
($0.1 - 0.5$ keV) and high energy bands ($0.5 - 2.0$ keV).  Better
definition of the pulse profile and spectrum in the energy range
1.0 - 30. keV by ASCA and RXTE have shown that, in the case of Vela and
Geminga, the high energy pulses are double-peaked and in phase with the $\gamma$-ray pulses measured by EGRET.  Furthermore, the 2.0 - 30 keV 
spectrum of Vela is consistent with an extrapolation of the OSSE spectrum
(Strickman et al. 1998).  The characteristics of the hard X-ray components 
in these pulsars are therefore best explained if their origin is 
non-thermal magnetospheric emission.  The phase shifts between the hard and
soft components seen by ROSAT may be understood as a transition from a
single broad thermal profile to a double-peaked non-thermal profile. 

The measured pulsed fractions and pulse widths are strongly energy 
dependent.  The pulsed fraction (defined in equation~[\ref{fp}]) of 
Geminga and PSR 0656+14 increases through 
the ROSAT band, starting at about $10-20\%$ around 0.1 keV and reaching
$80-90\%$ above 1 keV. But it is not clear how much of this 
increase is instrinsic
to the thermal component and how much is due to contamination by the
hard, non-thermal component, which is known to have a high pulsed fraction.
In the case of Geminga (Halpern \& Wang 1997), the pulse profile also changes 
significantly above
about 0.5 keV, where the power law spectral component becomes significant, 
indicating pulsed fraction contamination by the non-thermal component at the 
higher ROSAT energies.  We have therefore chosen to model the thermal
X-ray pulse profiles only in the lowest energy ROSAT bands available.
For the purposes of our modeling, we have taken the observed pulse
fractions of $20-30\%$ for Geminga, based on the observed profiles
given by Halpern \& Wang (1997) for the range 0.08 - 0.28 keV.
Table 1 summarizes the observed parameters of the X-ray and $\gamma$-ray
emission that are relevant to our modeling. 
The observed pulsed fractions for Vela and PSR 0656+14 are 11\% 
(\cite{oge93}) and 7\% (\cite{fin92}, \cite{fin97}, \cite{and93}), 
respectively. The pulse widths (FWHM -- Full Width at Half Maximum) in 
the lowest energy bands for all these pulsars lie roughly between 0.35 
and 0.5 of pulse phase.  We have determined approximate pulse widths, 
from the definition given in $\S$ 4, of $0.35 \pm 0.05$ for Geminga, 
$0.45 \pm 0.05$ for Vela, and $0.55 \pm 0.05$ for PSR0656+14.   

Vela and Geminga have 
well-defined high-energy $\gamma$-ray profiles, as seen by EGRET,
showing two sharp peaks with phase separations of 0.4 and 0.5 
respectively and emission between the peaks (\cite{Kan94}, \cite{MH94}). 
PSR 0656+14, however, is considered only as a possible detection of pulsed
emission by EGRET, with a poorly defined $\gamma$-ray profile (\cite{RM96}).  
For this
study, we will assume a double-peaked profile with phase separation of 0.4,
but we emphasize that this is highly uncertain.  The relative phases of the
X-ray and $\gamma$-ray profiles are reasonably well determined for Vela
and Geminga, but very uncertain for PSR 0656+14.  The absolute 
phase was not determined at the time of the ROSAT measurement 
(\cite{fin92}), but
we have taken in Figure 2 the relative phase between the EGRET and soft
ROSAT profiles as determined by Thompson (1997) from
comparison of EGRET, ASCA (both having absolute timing) and hard ROSAT 
profiles.  For all three pulsars, the broad X-ray profiles roughly
coincide in phase with the $\gamma$-ray profile.  The peaks in the X-ray
profile lie between the $\gamma$-ray peaks in the case of Geminga and Vela,
although both are offset by about 0.1 in phase toward the second $\gamma$-ray
peak.

\section{DESCRIPTION OF THE MODEL
         \label{sec:3}          }

In our modeling of the (soft) thermal X-ray emission from a NS
surface we include the effect of general-relativistic light bending 
which effectively smears out any intrinsic flux variation with pulsar 
rotational phase. Thus, the strong gravity of the NS tends to decrease 
the pulsed fraction, while the pulse width tends to increase 
(cf. \cite{pag95}). Much of the details of isotropic thermal X-ray 
emission from a NS surface 
have been discussed by Page (1995; and references therein), who 
concluded that the observed pulse fraction of Geminga's 
X-ray emission is difficult to explain unless the pulsar obliquity 
is close to $\pi /2$ and the surface magnetic field of the NS is 
highly non-dipolar. 
The available X- and $\gamma$-ray observational data on Geminga, 
Vela, and PSR 0656+14 summarized in \S~2 prompted us to explore 
the effect of X-ray beaming on the (supposedly) thermal soft 
X-ray emission from these pulsars.

It is generally expected (see discussion and references below) that the 
magnetic polar caps of a cooling NS are slightly hotter than the rest 
of the stellar surface, because of the effect of the strong magnetic 
field on heat transport in the surface layers of the NS. 
Hence, even an isotropically emitting NS surface may look somewhat 
brighter at the phases when the photons emitted 
from the polar caps get beamed into the observer's viewing angle. 
The presence of an intrinsic anisotropic component 
(e.g. associated with the anisotropy of radiative transfer in a 
strong magnetic field) may substantially enhance a contrast 
between fluxes received at the pole and the equator. The beaming of the 
X-ray emission along the magnetic field, as will be discussed 
in this paper, may allow a relatively large pulsed 
fraction even for a small obliquity and for a dipolar stellar 
magnetic field.

We assume that the $\gamma $-ray emission is generated above 
the pulsar polar cap within $\sim $ 1-3 stellar radii from the 
surface.  The mechanism for $\gamma$-ray production from the polar 
cap is a curvature-synchrotron/pair cascade from electrons 
accelerated in a region of open magnetic field lines above a  
polar cap (see \cite{dau96} for details). The efficiency of the  
electron-positron pair and $\gamma $-ray production 
varies across this region, so that in an axisymmetric case 
the emission is peaked in a hollow cone inscribed in the surface 
formed by the last open field lines.  The two narrow $\gamma $-ray 
peaks with interpeak emission observed in Geminga and Vela are well 
reproduced in such a ``hollow-cone" model.  

The calculation of the photon flux in a gravitational field of a 
slowly rotating NS is rather straightforward and has been 
described by many authors (see e.g. \cite{pec83}, \cite{rif88}, and 
\cite{pag95} for the most recent relevant discussions), and we shall 
not reproduce it in our paper. We must note however, that because 
of general-relativistic light bending the specific flux is now a first
moment of specific intensity with respect to $\cos \theta ^{\ast }$, 
where $\theta ^{\ast }$ is an angle between the photon wave vector 
and the local normal at the stellar surface, and the integration should 
be now taken over a solid angle ( $> 2 \pi $ steradian) determined by 
the maximum deflection angle (see e.g. \cite{pag95}) of a photon 
emitted nearly tangential to the surface. Note that in our 
calculations this angle (see \S\S~3.1 and 3.3), which is counted from 
the local normal at the emission point, is 131.9$^o$.

\subsection{Thermal X-Ray Emission
            \label{sec:3.1}}

We consider thermal black-body X-ray emission from the whole 
stellar surface, including the effects of general-relativistic 
light bending and anisotropization of emission produced by a 
dipolar surface magnetic field. 

It is important that the effect of a strong magnetic field on 
the X-ray emission is twofold:

\begin{enumerate}

\item anisotropy of the surface temperature distribution: 
the transverse heat conductivity is suppressed due to magnetization 
of electrons, which results in a surface temperature at 
the magnetic pole that is e.g. a factor of 1.5-2.5 higher than 
at the magnetic equator; 

\item beaming of the thermal emission along a tangent to the 
magnetic field lines in a strongly magnetized
atmosphere (when $\rm h\nu _c \gg kT$, where $\rm \nu _c = 
eB/2\pi m_ec$ is the electron cyclotron frequency in a magnetic 
field of strength B). 

\end{enumerate}

The effect of beaming of the emission results from the angle and
polarization dependent
opacity in a NS atmosphere, that is lower along the magnetic field than
in the transverse direction. The reduction of the photon opacity in 
the magnetic field of a NS was first discussed by Cohen et al. 
(1970) and Tsuruta et al. (1972), and addressed in 
more detail by Lodenquai et al. (1974). These authors found that 
(see \cite{lcrt74}) in a magnetized plasma two independent modes 
(ordinary and extraordinary modes) of an electromagnetic wave have 
significantly different mean free paths. For example, the 
extraordinary-mode photons (for which the electric field wavevector 
is perpendicular to the magnetic field) generally have a 
much longer mean free path than ordinary-mode photons and give the main 
contribution to radiation transport in the very surface layers of a
NS atmosphere. Lodenquai et al. (1974) have presented the approximate 
relation between scattering cross-sections 
with and without strong magnetic field (e.g. $\gsim 10^{12}$ G) 
for the extraordinary mode: 
\be
\rm \sigma _{\perp }(B) \approx \left( {{\omega }\over 
{\omega _c}} \right) ^2 \sigma (0) , ~~~~~~~~~~~~~~~~
\omega \ll \omega _c \equiv 2\pi \nu _c .
\ee
where $\omega $ is the frequency of a photon.

Many authors have since calculated the radiative thermal
conductivities in a plasma with a magnetic field (see e.g. \cite{pav77}, 
\cite{sil80}, and references therein) and have discussed in 
more details the effects of opacity on the emerging emission from 
a magnetized atmosphere of a NS (see e.g. \cite{kam82}, and
\cite{pav94}). 

The anisotropy of the surface temperature distribution due to 
anisotropy of electron thermal conduction in a strong 
magnetic field ($\sim 10^{12}-10^{13}$ G) in a NS crust can be  
adequately described by an approximate formula (see \cite{gre83}): 
\be
\rm T_*(\Theta ) = T_p (\cos ^2 \Theta + \chi _0 ^4 \sin ^2 \Theta )^{1/4}, 
\ee
where $\rm T_p$ is the effective temperature at the magnetic
pole, $\Theta $ is the angle between the local normal to the surface 
and tangent to the field line, $\rm \chi _0 = T_{eq}/T_p \approx
0.3-0.6$, $\rm T_{eq}$ is the effective temperature at the 
magnetic equator. The parameter $\chi _0$ can be expressed in 
terms of the physical quantities as $\rm \chi _0 = 
(\kappa _{\perp }/\kappa _{\parallel })^{1/4}$, where $\kappa
_{\parallel }$ and $\kappa _{\perp }$ are the thermal conductivities
in the surface layers of a NS along and perpendicular to the direction
of a magnetic field, respectively.  The values of 
$\chi _0$ can be estimated 
from the dependences between the surface and internal temperatures 
of a cooling NS calculated for the range of magnetic fields 
$\leq 10^{10}-10^{14}$ G and presented by Page (1995, Figure 1). 
However, the results of our calculations are rather insensitive to the 
particular value of $\chi _0$ we choose from the range specified 
above. The effect of anisotropic cooling and atmospheric radiation of 
neutron stars with strong magnetic fields have been discussed by
Shibanov et al. (1995, see also references therein). In our
calculation we assume a dipole magnetic field modified by the static 
part of the gravitational field (see e.g. \cite{mus87}). 

In our analysis, to incorporate the effect of X-ray beaming, we 
exploit the results of the numerical calculations of radiative
transfer from a magnetized atmosphere of a NS (Zavlin et al. 1995) which 
indicate that the emission pattern consists of a peaked (along the 
direction of the magnetic field) pencil beam component and a broad, 
nearly isotropic, fan beam component. In Figure 3 we present the normalized 
profiles of the function $\rm I_{\nu }(\theta _B) \cos(\theta _B)$ 
we use in our calculations, where $\rm I_{\nu }$ is the specific 
intensity of radiation, and $\rm \theta _B$ is the angle between the 
wave vector and tangent to a magnetic field line at the stellar surface. 
These profiles represent two examples having beamed components of
different shapes: narrow and weak, and broad and strong. 
They correspond to the relatively low ($\sim 0.18$ keV) and high 
($\sim 1$ keV) photon energies and are very similar to those 
provided by Zavlin, Shibanov \& Pavlov (1995) in Figures 2a and 2b, 
respectively (that are calculated for the stellar effective
temperature of $3\times 10^{6}$ K and surface value of the magnetic
field strength of $\sim 10^{12}$ G). We must note 
that these angular profiles are generally energy dependent and for
different parameters may look different. For example, the relative
magnitude of the beamed component increases with photon energy and
also with an increase in surface gravity. Qualitatively, as has
been discussed by Zavlin, Shibanov \& Pavlov (1995), the beamed
component which is determined by the enhanced atmospheric transparency
along the magnetic field, corresponds to the photons emerging from the 
deeper and hotter layers along the field, and the dependence on the
surface gravity results from the emerging radiation dependence on the
temperature scale height.

We calculate the X-ray flux received by a distant observer 
using the following expression (cf. \cite{pag95})
\be
\rm F_{\nu }(E,T, \alpha , \zeta ) = F_0 
\int _0^{\theta _{max}} \int _0^{2\pi } I_{\nu} 
(E_{\ast } ,T_{\ast }, \theta _B)
\cos [\theta ^{\ast }(\theta )] \sin \theta d\theta d\phi ,
\ee
where E and T are the red-shifted 
energy of a photon and stellar effective temperature, respectively,  
$\rm E_{\ast }$ and $\rm T_{\ast }$ are the photon energy and stellar 
effective temperature as measured at the stellar
surface, $\alpha $ and $\zeta $ are the obliquity and observer angle, 
respectively, $\rm F_0$ is a normalization constant, 
$\theta $ and $\phi $ are the angles of a spherical coordinate system 
with the axis along the local normal at the point of photon emission, 
$\theta _{\rm max}=131.9^o$ is the maximum 
angle between the normal at the point of photon emission and the wave 
vector of a photon received by the observer, $\theta ^{\ast }(\theta )$ is
the angle between the photon wave vector and the local normal at the
stellar surface. The relation between the angle $\theta ^{\ast }$ and 
angle $\theta $, at which the photon reaches the observer, is defined
by (see e.g. \cite{pec83} for details) 

\be
\rm \theta (\theta ^{\ast }) = {\it a} \int _0^{\varepsilon } 
{{d{\it x}} \over {\sqrt{1-(1-{\it x}){\it x}^2{\it a}^2}}} ,
\ee
where $\rm {\it a} = \sin \theta ^{\ast }/[\varepsilon 
(1-\varepsilon )^{1/2}]$, $\rm \varepsilon = r_g/R$ and $\rm r_g = 
2GM/c^2$ is the gravitational radius of a NS of mass M and 
radius R.

\subsection{Non-thermal Gamma-Ray Emission
            \label{sec:3.2}}

Most of the observed $\gamma$-ray pulsars have double-peaked 
profiles with peak phase separations of 0.4 - 0.5 (\cite{tho96}). 
There are presently two types of pulsar $\gamma$-ray emission models
which have been studied in detail.  Polar cap models consider 
electrostatic acceleration above the neutron star surface near the
magnetic poles, arising from charge starvation caused by field line
curvature (Arons 1983) and general relativistic inertial frame
dragging (Muslimov \& Tsygan 1992, Muslimov \& Harding 1997).
The $\gamma $-ray emission in polar cap models (\cite{dau94}, 
\cite{dau96}; \cite{stu94}) is a hollow cone, with opening angle 
$\theta_{\gamma}$, centered on the magnetic pole (see Figure 1), 
producing either double-peaked or single-peaked pulse profiles
depending on observer orientation. 
Outer gap models consider acceleration in the outer magnetosphere
in ``Holloway" gaps caused by charge depletion along the null charge
surfaces (Cheng, Ho \& Ruderman 1986).  The $\gamma$-ray emission
pattern is a curved fan beam described by the last open magnetic field
line (Romani \& Yadigaroglu 1995).
Both models can produce double-peaked $\gamma$-ray profiles simlilar 
to what is
observed, but the phase of the magnetic pole relative to the two 
$\gamma$-ray peaks is very different in the two types of model.
For the polar cap models, the phase of closest approach to the
magnetic pole, and thus the peak of the thermal X-ray pulse, lies 
midway between the $\gamma$-ray peaks, in the
interpeak emission region.  In the outer gap models, the predicted 
phase of closest approach to the magnetic pole and the thermal X-ray 
pulse lies outside the $\gamma$-ray peaks.  

In this paper, we consider the hollow-cone $\gamma$-ray beam pattern
predicted by polar cap models.
When $\theta_{\gamma} \sim \alpha$,
an observer may see a broad double-peaked $\gamma$-ray pulse profile 
with the peak separation $\rm \Delta \phi $ given by (\cite{dau94})
\be  \label{phi}
\rm \cos (\Delta \phi ) =
{{\cos\theta_{\gamma} - \cos\alpha\cos\zeta }
\over {\sin \alpha \sin \zeta }},
\ee
where $\theta_{\gamma}$ is the $\gamma$-ray beam opening angle. The 
$\gamma$-ray beam opening angle is determined approximately by the 
locus of the tangent to the outermost open field line:
\be  \label{thetagam}
\rm \tan \theta_{\gamma } \simeq {{3\theta_{pc}(1-\theta_{pc}^2r/R)^{1/2}
(r/R)^{1/2}} \over {3(1-\theta_{pc}^2r/R) - 1}} = 
{3\over 2}\left( {{\Omega r}\over c} \right) ^{1/2} 
{{(1-r/R_{LC})^{1/2}}\over {(1-3r/2R_{LC})}}, 
\ee
where $\rm \theta_{pc}$ is the polar cap half-angle, r is the 
radius of emission, R is the stellar radius, $\Omega $ is the angular 
velocity of stellar rotation, and $\rm R_{LC}\equiv c/\Omega $ is 
the light-cylinder radius. 
In this formula the equality holds for the standard value of the 
polar cap half-angle, $\rm \theta_{pc} = (\Omega R/c)^{1/2}$. 
If $\rm r > R$ and/or the polar cap half-angle is larger than the
standard value, then 
$\theta_{\gamma }$ could be as large as $20^0 - 30^0$ (Figure 1). General 
relativistic effects on the photon trajectory and on the dipole 
magnetic field structure introduce small corrections to 
$\theta_{\gamma}$ and $\rm \theta_{pc}$.

For an observed phase separation $\rm \Delta \phi $ between the first and 
second peaks of the $\gamma$-ray pulse profile, equation~(\ref{phi}) 
describes closed contours in the $\zeta-\alpha$ plane, as a function of 
$\theta_{\gamma}$.  Such contours for Vela ($\rm \Delta \phi = 0.4$) and
Geminga($\rm \Delta \phi = 0.5$) are shown as dashed lines in 
Figures 6 and 7, for values of $\theta_{\gamma}$ between $5^0$ and 
$35^0$.  In the case of $\rm \Delta \phi = 0.5$ the contours collapse 
to single lines, so that the relation between $\zeta$ and $\alpha$ 
is single-valued. The predicted $\gamma$-ray beam opening angles 
$\theta_{\gamma}$ (see e.g. \cite{dau96}) from 
equation~(\ref{thetagam}) are between $\sim 2^0 - 20^0$ for Geminga 
($\Omega = 26.5\,\rm s^{-1}$) and between $\sim 3^0 - 35^0$ for Vela 
($\Omega = 70.6\,\rm s^{-1}$), for $\rm r = (1 - 3)R$ and 
$\rm \theta_{pc}$ up to 5 times the standard value.  Therefore, 
small values of both $\zeta$ and $\alpha$ ($< 35^0$) are favored 
in polar cap models.
            
\subsection{Main Input Parameters
            \label{sec:3.3}}

We use a canonical NS model of mass $\rm 1.4~M_{\odot }$ and 
radius of 10 km. Note that for this stellar model a photon emitted 
almost tangential to the stellar surface gets deflected (while 
remaining in the same plane with a tangent to the trajectory and the 
normal to the stellar surface at the point of emission) by an 
angle of $131.9^0$ to the normal. We must point out that the canonical 
NS model is consistent with the recent RXTE results by the Illinois 
group (Miller, Lamb \& Psaltis 1997) which constrain NS equation of 
state (admittedly for NSs in Low-Mass X-ray Binaries rather than for 
isolated NSs).

In Figure 3 we show two characteristic emission patterns for the beamed 
surface X-ray emission we have used in our calculations. In general, 
the intensity of this beamed emission can be presented as a linear 
combination of the pencil and fan components. In this paper 
we have chosen a particular sample (see Figure 3) consistent with 
the results of a modeling of emergent spectra from a magnetized 
hydrogen atmosphere of a NS (see e.g. \cite{zav95}). We must note that
there is a qualitative difference between the profiles shown in Figure
3, that should manifest itself in the resulting pattern of contours of
constant X-ray pulsed fraction in the $\alpha$-$\zeta $ map (see
\S~4). The profile shown in Figure 3 by a
solid line (case 1) has a substantial pencil component (of angular 
half-width of $\sim 15^o$) superposed on a relatively weak fan
component, while the profile shown by a dotted line (case 2) 
has a very narrow (of angular half-width of $\sim 3^o$) pencil beam on 
top of a larger fan component. For case 1, the contribution 
of the beamed component to the integrated X-ray flux is substantial, and 
one can expect that the X-ray pulsed fraction and pulse width in this 
case will be determined by the beamed component. On the contrary, 
for case 2 the contribution of the beamed component is rather
small, and both the X-ray pulsed fraction and pulse width will be 
mostly determined by the fan component. In our calculations we 
assume that the surface effective temperature (at the magnetic
pole) and polar value of the magnetic field strength 
are respectively, $5\times 10^{5}$ K (Geminga), $10^6$ K (Vela 
and PSR 0656+14) and $\sim 3 \times 10^{12}$ G (Geminga), 
$\sim 6 \times 10^{12}$ G (Vela and PSR 0656+14). For both these cases 
in our calculations we used the value $\chi _0 = 0.3$.

As has been noted by Page (1995), the response 
of the ROSAT PSPC results in a distortion of the observed pulsed 
fractions due to the mixing of photons with 
different energies in the detector.  Photons of different energy also
suffer varying degrees of interstellar absorption.
The result of these effects is an increase in the observed
pulsed fraction  (up to $\sim 70-80$ \%) at lower energies. 
In our modeling we have not included a detector response funtion 
(we assume a
``perfect" detector) or intersellar absorption effects, so that the 
calculated pulsed fractions (at the energy of 0.18 keV) may be 
systematically lower than those calculated including these effects.

\section{RESULTS OF NUMERICAL CALCULATION
         \label{sec:4}                             }

In our calculations we use the following more or less generally 
accepted definition of pulsed fraction, that is sometimes also referred to
as the modulation index (for a given photon energy):
\be \label{fp}
\rm f_{pulsed} = {{F_{max}-F_{min}}\over {F_{max}+F_{min}}} ,
\ee
where $\rm F_{min}$ and $\rm F_{max}$ are the minimum and maximum 
values of the photon flux. We define the pulse width (or FWHM) as the
phase difference between the right and left wing of the pulse at 
half maximum (at $\rm F_{max}/2$). 

In Figure 4 we illustrate our calculations of the X-ray pulsed 
fraction (in \%) (left panel) and pulse width (right panel) for 
the case of isotropic emission. In our calculations we have assumed 
the stellar effective temperature (at the magnetic pole) of $5\times
10^5$ K. The results of our calculations are in reasonable agreement
with those published earlier by other authors. Figure 4 shows that for
the range of X-ray pulse widths of 0.35-0.5 (matching the observed
ones) the pulsed fraction (at the energy of 0.18 keV) hardly exceeds a
few percent. Also, Figure 4 indicates that the pulse width decreases 
as the pulsed fraction increases. Perhaps this tendency was one of the
motivations for modeling Geminga's X-ray emission (see \cite{pag95}) 
in terms of the orthogonal rotator with highly non-dipolar magnetic 
field, because the former favors largest pulsed fractions while 
the latter may broaden the pulse(for an appropriately chosen combination 
of magnetic multipoles).

The principal result of our calculations is that the 
observed thermal X-ray light curves and profiles can be produced
by an anisotropic emission pattern even for small obliquity angles. 
In Figure 5 we present the X-ray light curves calculated for the
emission patterns with different contributions from the beamed 
component. The left panel in Figure 5 illustrates the 
light curves calculated for the emission pattern shown in Figure 3 
by a solid line (case 1) for different angles $\alpha $ and $\zeta $
(solid line: $\alpha = \zeta = 10^0$, dotted line: 
$\alpha = 15^0$, $\zeta = 50^0$, and dashed line: 
$\alpha = \zeta = 90^0$). The right panel in Figure 5 illustrates the 
light curves calculated for the fixed angles $\alpha = \zeta = 15^0$
and for three different cases of emission pattern where, respectively,
the beamed component dominates (solid line), beamed component is 
suppressed (dotted line), and where the contributions from the beamed 
and fan components are that given in Figure 3 by a solid line.

Figures 6 - 9 summarize the results of our modeling of the soft 
X-ray and hard $\gamma$-ray emission for Geminga and Vela (the results 
obtained for Vela also apply to PSR 0656+14), respectively. Figures 
6 and 8 correspond to the case of the X-ray emission pattern shown
in Figure 3 by a solid line (case 1), while Figures 7 and 9
correspond to the case-2 emission pattern (Figure 3, dotted line). The
calculations presented in Figures 6, 7 and 8, 9 have been 
performed for the effective stellar temperatures (at the magnetic
pole) of $5\times 10^5$ and $10^6$ K, respectively. In Figures 6 - 9 
the left and right panels display the calculated X-ray pulsed 
fractions (\%) and pulse widths, respectively. The shaded areas 
in the left panels of Figure 6, 8 and 9 denote 
the range of the observed pulsed fractions. Our modeling of Geminga's 
X-ray emission (Figures 6 and 7) shows that the pulsed fractions 
calculated for the
emission pattern shown in Figure 3 by a dotted line (case 2) are well 
below the observed ones. Thus, at least for Geminga, the surface X-ray
emission should have a rather strongly beamed component.

Note that for the case-1 X-ray emission pattern (Figure 3, solid line) 
and for $\alpha \sim \zeta$ the pulse width gets smaller than 0.3 at 
$\alpha \sim \zeta \gsim 20^0$, while for the case-2 emission 
pattern this already occurs at $\alpha \sim \zeta \gsim 10^0$. 
This can be easily understood in terms of different angular 
widths of the beamed component of the X-ray emission patterns shown 
in Figure 3. Also, Figures 6 - 9 show that the curvature of contours 
of equal X-ray pulsed fraction and pulse width is determined by the 
angular width of the beamed component, whereas the values of the 
X-ray pulsed fraction and pulse width are determined by the relative
contribution of the beamed component (besides a general contribution 
determined by the input parameters such as the effective temperature, 
photon energy, stellar mass and radius, etc).

Our calculations show that if one of the angles 
$\alpha $ and $\zeta $ is $\gsim 15-20^o$, then the contours 
labelling the observed values of pulse width (0.35 - 0.5) have 
no overlap with 
those labelling pulsed fraction of $>$ 10-20 \% . This means 
that our modeling of the X-ray pulsed emission alone already 
constrains the angles $\alpha $ and $\zeta $ to the small 
values of $\lsim 30^0$. This tendency gets more pronounced for the case 
of a sharp pencil beam component (see Figures 7 and 9, right panels). 
In this case the contours of constant pulsed fraction (of $\gsim 
10 \% $) degenerate into very elongated parabolas symmetric about the
diagonal $\zeta = \alpha $ and with the vertices in the low-left 
corner. This implies that the relatively high observed pulsed fractions 
(of order of 10-20\% ) and large pulse widths (of order of 
0.35 -0.5) may be allowed only for the relatively small obliquity and 
observer angles, $\alpha , \zeta \lsim 10^o$.  Thus, the main
qualitative result of our modeling of the soft thermal X-ray 
emission from Geminga, Vela, and PSR 0656+14 is that the presence 
of a pencil component in emission is necessary and constrains the 
observer and obliquity angles to small values, and that the 
sharper a pencil component is, the closer these angles are to 
each other. 

Figures 6 and 7 show that only the broad beam pattern (case 1) will
produce a pulse fraction high enough to account for Geminga's
observed pulse fraction.  The pulse width contours in Figure 6
matching the observed range of pulse widths ($0.35 \pm 0.05$) also
overlaps the allowed phase space of pulse fraction.  Figure 8 and 9
show that there is allowed phase space of pulse fraction for both
beaming paterns for Vela and PSR 0656+14, but the larger observed 
pulse widths of these pulsars is consistent with only the broad beam
pattern (case 1) in Figure 8.  A larger pencil beam component therefore
seems to be favored for all three sources.

Assuming values of the opening angle 
for the hollow-cone $\gamma $-ray emission allows us to 
quantitatively and independently estimate the angles $\alpha $ 
and $\zeta $ for these pulsars. 
The contours of constant $\alpha $ and $\zeta $ in the hollow-cone
emission model for the observed
values of $\Delta\phi$ for these pulsars, shown by dashed lines 
in Figures 6 - 9,
constrains both $\alpha $ and $\zeta $ to be small. For particular model parameters we employ in this paper these angles should be 
$\lsim 20^0-30^0$.  This constraint
from modeling the $\gamma$-ray pulse profiles is thus in agreement
with those derived from modeling of the X-ray profiles.  From the
combined requirements of large X-ray pulsed fraction ($\rm f_{pulsed} > 
10\% - 30\%$), X-ray pulse width (0.35 - 0.5) and $\gamma$-ray pulse
separation, the $\gamma$-ray beam hollow-cone opening angles must
be $13^0 \lsim \theta_{\gamma} \lsim 30^0$ for Geminga and
$5^0 \lsim \theta_{\gamma} \lsim 30^0$ for Vela.

If the pulsed thermal soft X-ray emission from Geminga, Vela, and 
PSR 0656+14 is dominated by a beamed component (produced e.g. 
by the effect of anisotropic opacity of a magnetized atmosphere), 
then the model X-ray light curves for these pulsars agree very well 
with the observed pulsed fractions and pulse widths of their 
soft X-ray emission (at a median energy of 0.18 keV). The 
calculated $\gamma $-ray profiles match the 
observed pulse spacing, $\Delta \phi $, for these pulsars. The 
simultaneous modeling of the soft (thermal) X-ray and 
$\gamma $-ray emission for Geminga, Vela, and PSR 0656+14 constrains 
the phase space for the possible obliquity and observer angles, 
and favors relatively small values for these angles (Figures 6 - 9).

\section{SUMMARY
         \label{sec:5}}

We have discussed the issue of whether the observed $\gamma $-ray 
emission and recently detected soft (presumably thermal) X-ray 
emission from Geminga, Vela and PSR 0656+14 can be understood in terms
of a polar cap (or hollow cone) model proposed for the $\gamma $-ray
emission from pulsars.  We have calculated the range of observer and 
obliquity angles allowed by the observed pulsed fractions and widths of
the soft X-ray profiles of these pulsars (at the median energy of 
$\sim $0.18 keV), assuming anisotropy of the X-ray emission in a
magnetized atmosphere.  We found that small values of observer angle
and obliquity are required to account for the single, 
relatively broad (with a phase width of $\sim $0.35-0.5) X-ray peaks
and these values can still produce the observed pulsed fractions of 
the X-ray profiles.  The range of these angles restricted by the X-ray 
profiles are found to be consistent with those values required to
reproduce the observed narrow double $\gamma $-ray 
peaks separated by a phase interval of 0.4-0.5.  In addition, the
appearance of a single broad X-ray pulse between the two $\gamma$-ray
peaks predicted by polar cap models seems to be borne out at least
for Vela and Geminga. 

Our main results can be summarized as follows.

\begin{enumerate}

\item The possibility of beaming of the thermal X-ray emission in Geminga,
Vela, and PSR 0656+14 provides a consistent explanation for their
observed X-ray light curves and is in accord with the polar cap models
for their $\gamma $-ray emission.

\item The obliquity and/or observer angle in Geminga, Vela, 
and PSR 0656+14 may be less than 30$^0$. The $\gamma$-ray opening
angles must be at least 13$^0$ for Geminga and at least 5$^0$ for 
Vela and PSR 0656+14.

\item For anisotropic X-ray emission, the observed pulsed fraction 
and pulse width are much less sensitive to the effective temperature 
and are determined primarily by the degree of a beaming.

\item The magnitude of X-ray pulsed fraction is mainly determined 
by the magnitude of the beamed emission relative to the fan beam
emission. In the case of a very strong beamed 
component, whose contribution to the X-ray flux is significant, 
the maximum pulsed fractions 
should be observed for a rather wide range of obliquity and observer 
angles. 

\item The range of observer and obliquity angles allowing for 
the largest possible pulsed fraction is determined by
the angular width of the beamed component.
In the case of a very narrow beamed component, the maximum 
pulsed fractions will be observed when the obliquity and observer 
angles are very close to each other. 

\end{enumerate}

Recently, Tauris \& Manchester (1998) have reanalyzed radio pulsar 
polarization data to compute the obliquity distribution of the parent 
population of all radio pulsars.  Their derived distribution peaks at
small obliquities and suggests that most pulsars have $\alpha \lsim 
35^0$.  Our results are thus consistent with this picture.  

There are several important issues that we have thus far not 
addressed in our modeling of X-ray and $\gamma$-ray pulse profles:
polar cap heating and plausible physical 
reasons for the observed phase offset between X-ray and $\gamma-$ray 
pulses.  Since the pioneering study by \cite{rud75} it has
been understood that the development of electron-positron cascades 
above the NS surface initiated by primary electrons 
should unavoidably result in precipitation of ultra-relativistic 
positrons on the stellar surface. The kinetic energy of these positrons 
should be eventually transformed into the thermal energy of the NS 
crust and then reradiated most likely in soft X-rays (see e.g. 
Arons \& Scharlemann 1979, and Arons 1981). The polar cap heating could
thus add a component to the thermal X-ray pulse profiles. 
The efficiency of the
polar cap heating depends on the number density of positrons that 
flow back from the pair formation front (PFF) to the stellar 
surface. This number density cannot be calculated from first 
principles. Instead it is rather sensitively determined by the 
transverse and longitudinal structure of the PFF, electrodynamics 
of the PFF, and by the dynamics of positron acceleration (Harding \& 
Muslimov 1997b).
Estimates of X-ray luminosity due to polar cap heating by Arons (1981)
predict that such heating accounts for only $8\%$, $0.12\%$ and 
$0.005\%$ of the observed thermal X-ray luminosity of Vela, Geminga
and PSR 0656+14, respectively (Harding 1995).  

However, these polar
cap heating estimates need to be revised using more recent calculations
of the electric fields above the polar cap (Muslimov \& Tsygan 1990, 1992, 
and Muslimov \& Harding 1997).
We can make the 
following rough estimates based on the results of the general relativistic
treatment of the acceleration of the primary beam over the polar cap.
The main difference between this
and the classical treatment (see also Arons 1996) is that 
the general-relativistic dragging of inertial frames allows very efficient 
acceleration even for an aligned rotator and does not require the 
concept of  ``favorably curved field lines" introduced by \cite{aro79}. 
In the regime of the space-charge-limitation 
of current, the electric field in the region of electron acceleration 
is determined by the difference between the local charge density 
of electrons and Goldreich-Julian charge density 
$\rm \Delta \rho _e$. This difference reaches a maximum 
$\rm |\Delta \rho _e|_{max} \sim |\rm \rho _{GJ}| \kappa $ 
(where $\rm \kappa \approx 0.38 r_g/R \sim 0.15$, $\rm r_g$ 
is the stellar gravitational radius,
see e.g. Muslimov \& Harding 1997) at some height $\rm h_m$ 
above the surface and then exponentially declines toward the PFF. 
The backflowing positrons 
enter a regime of relativistic motion near $\rm h_m$, where 
$\rm |\Delta \rho _e| \sim |\Delta \rho _e|_{max}$. The 
backflowing positrons tend to reduce the difference 
$\rm |\Delta \rho _e|$ and therefore the 
maximum value of the electric field. Thus, the maximum charge density 
of backflowing positrons can be estimated as $\rm |\Delta \rho _p| 
\sim |\Delta \rho _e|_{max}$. The total power put in
the acceleration of these positrons can be estimated as (see also 
Muslimov \& Harding 1997) $\rm L^{+}\sim f~L^{-}_{max}$, where $\rm 
L^{-}_{max}$ is the maximum power of primaries, and $\rm f \approx
0.25 \kappa $. Since $\rm L^{-}_{max}\sim L_{\gamma}$, we get 
that $\rm L^{+}\sim f~L_{\gamma}$. For a canonical NS of
mass 1.4 $\rm M_{\odot }$ and radius 10 km we get the values 
$\rm L^{+}$ that may account in part for the observed soft X-ray
fluxes from pulsars (see e.g. Becker \& Tr\"umper 1997). 

We now discuss the possible explanation for the phase offset 
(by about 0.15 in phase) of the X-ray pulse center  
toward the trailing $\gamma$-ray peak, observed in 
Geminga and Vela (see \S~2). In the framework of the polar cap 
model, the phase offset might be explained in terms of e.g. 1) an offset 
dipole  2) asymmetric polar cap heating  and/or 3) dragging of photon 
geodesics by the gravitational field of a rotating body (see e.g. 
Misner, Thorne, \& Wheeler 1973). 
For the offset dipole, the geometrical vertex of the
$\gamma-$ray emitting cone projects onto the stellar surface at a point
offset from the 
magnetic pole. This means that the phase of the 
thermal X-ray emission (centered at the magnetic pole) will be offset
from the $\gamma-$ray pulses.  A significant polar cap heating component
may add to the thermal X-ray profile caused by cooling.  If the heating
by precipitating particles is not uniform over the polar cap, then the
resulting X-ray component may add more at the trailing edge of the 
thermal pulse.  We have not yet made a detailed calculation of the 
distribution of heating rates over the polar cap, so this effect is
hard to predict at present.  Any offset due to asymmetric particle 
heating might be enhanced by an asymmetric $\gamma-$ray precipitation 
on the stellar surface near the polar cap from downward cascades due to
positron acceleration. The effect of frame dragging on the
light rays results in a phase shift of both X-ray and $\gamma-$ray 
pulses. However, since the 
magnitude of this effect has not been accurately evaluated and is
beyond the scope of this paper, we cannot say whether this effect will 
quantitatively account for the 
observed phase offset for a 0.1-0.3 s pulsar. 

In forthcoming publications we plan to discuss these 
and other effects in more detail and present more 
comprehensive theoretical analysis of the observed X- and 
$\gamma $-ray emission from Geminga, Vela, PSR 0656+14, and 
other pulsars from this subpopulation of radio pulsars.

\acknowledgments 

We thank George Pavlov for discussion, and Dany Page and Slava Zavlin 
for providing a sample of data to test our numerical code. 
We are also grateful to Dave Thompson for providing us with his
compilation of observed $\gamma$-ray and X-ray pulse profiles. A.K.H. 
acknowledges support through a NASA Astrophysics Theory Grant. 


\clearpage
 
\begin{deluxetable}{crrrrrr}
\footnotesize
\tablecaption{Observed Characteristics and Model Parameters \label{tbl-1}}
\tablewidth{0pt}
\tablehead{
\colhead{~~~} & \multicolumn{4}{c}{Observations} & 
\multicolumn{2}{c}{Model} \\
\colhead{Pulsar} & \colhead{$\rm B_p$}   & \colhead{$\rm T$}   & 
\colhead{$\rm f_{pulsed}$} & \colhead{$\Delta \phi$} & 
\colhead{$\rm T_p$} & \colhead{$\rm B_p$} \\
\colhead{~~~} & \colhead{($\rm 10^{12}$ G)}& \colhead{($10^5$ K)}  
& \colhead{~~~} & \colhead{~~~} 
& \colhead{($10^5$ K)} & \colhead{$\rm 10^{12}$ G}
} 
\startdata
Geminga & 3.3 & $5.6\pm 0.6$ \tnm{a} & 20-30\% 
& 0.5 & 5 & 3 \nl
Vela & 6.7 & 15-16 \tnm{b} & 11\%  
& 0.4 & 10 & 6 \nl
PSR 0656+14 & 9.3  & 8 \tnm{c} & 7\% \tnm{d} 
& ? & 10 & 6 \nl
\enddata
 
\tnt{a}{\cite{hal97}}
\tnt{b}{\cite{oge93}}
\tnt{c}{\cite{gre96}}
\tnt{d}{\cite{fin92}; \cite{and93}}
\end{deluxetable}

\clearpage
\vskip -2.0in
\psfig{figure=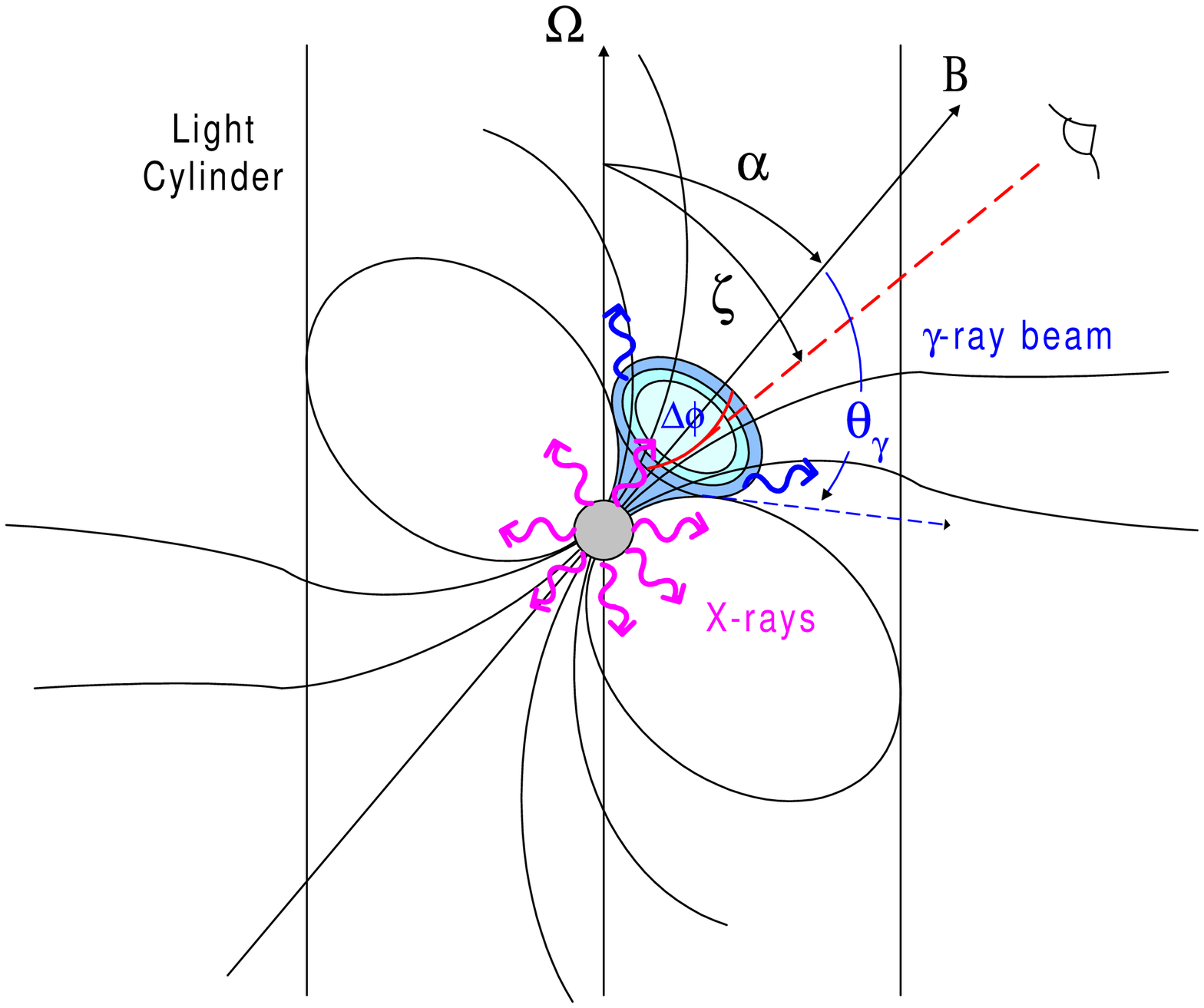,width=5.6in,angle=0} 
 \figcaption{ 
Schematic illustration of a polar cap $\gamma $-ray beam
geometry. Here $\pmb{$\Omega $}$ and $\pmb{$\mu $}$ are the vectors 
of stellar angular velocity and magnetic moment, respectively. 
The angles $\alpha $, $\zeta $, and $\theta _{\gamma }$ specify the 
obliquity, observer angle, and half-opening angle of a $\gamma $-ray
emitting hollow cone, respectively. The phase separation between the 
$\gamma $-ray pulses is $\Delta \phi $. 
\label{fig:1}}

\figureout{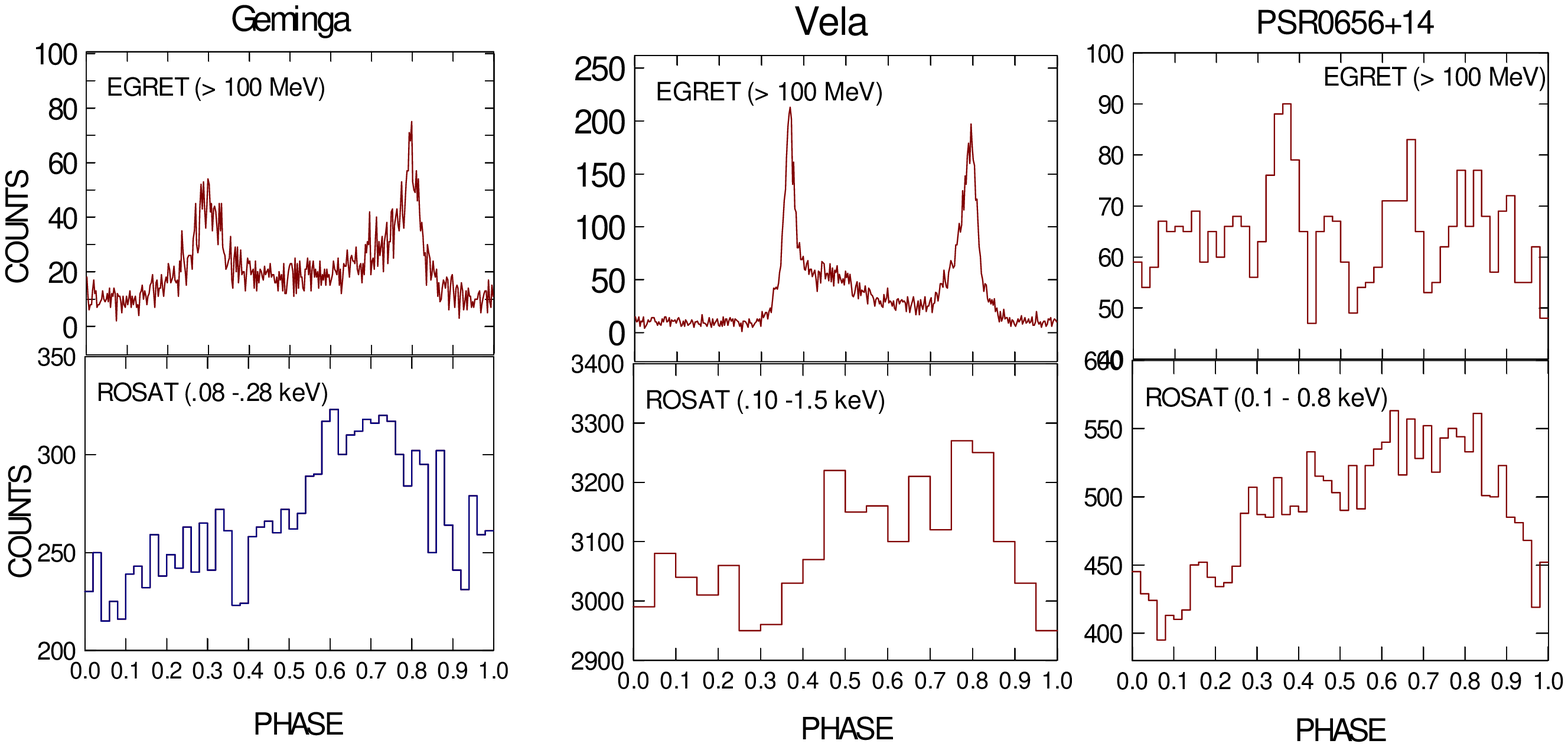}{0}{
Observed X-ray (lower plots) and $\gamma $-ray (upper plots)
light curves for Geminga (a), Vela (b), and PSR 0656+14 (c). See 
\S~2 for details.
\label{fig:2}}

\figureout{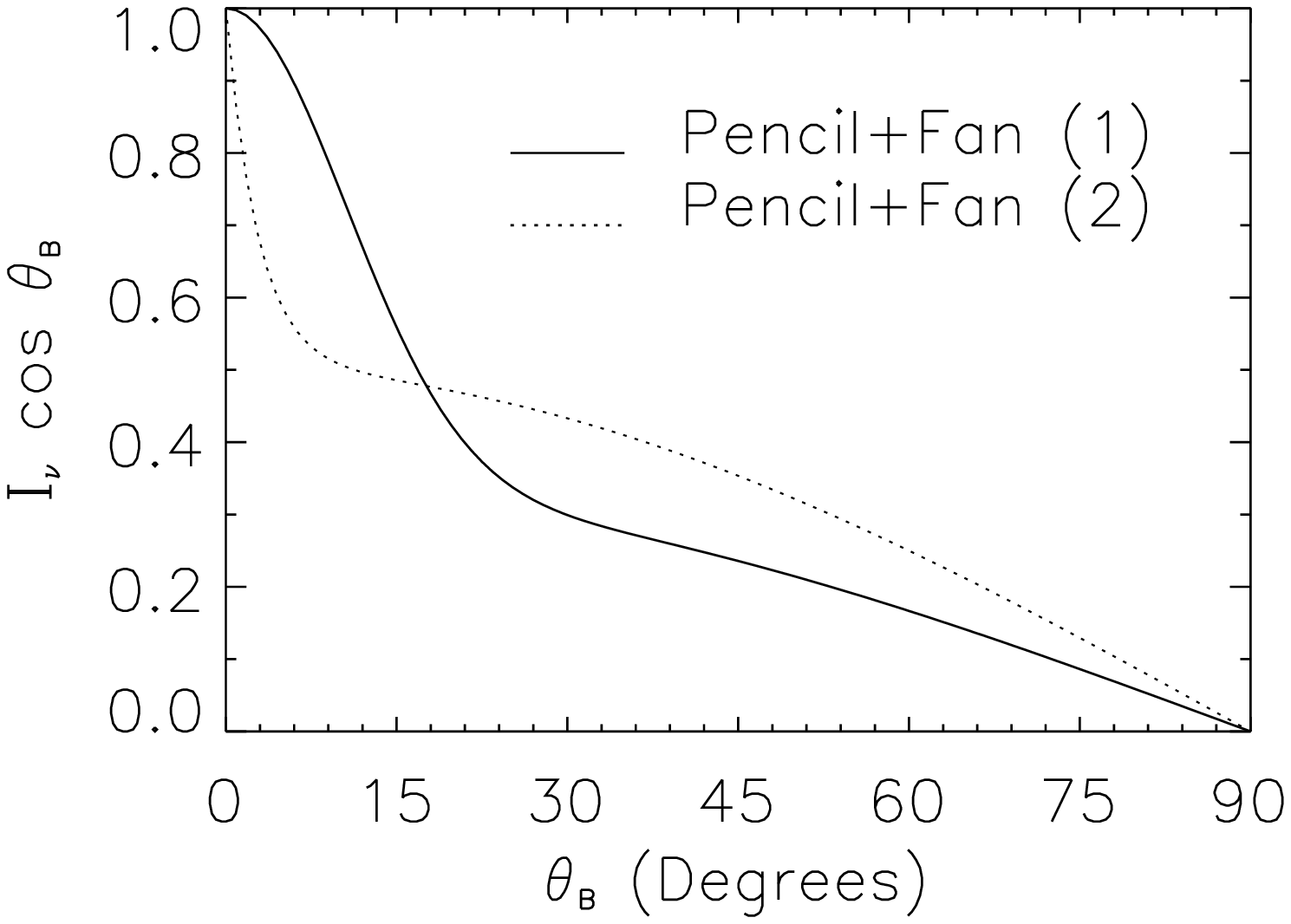}{0}{
Two different profiles assumed for the anisotropic
distribution of thermal X-ray emission (see \S~3.1 for details). 
Here $\rm \theta _B$ is an angle between the wave vector and 
tangent to the magnetic field line at the stellar surface.   
\label{fig:3}}

\figureout{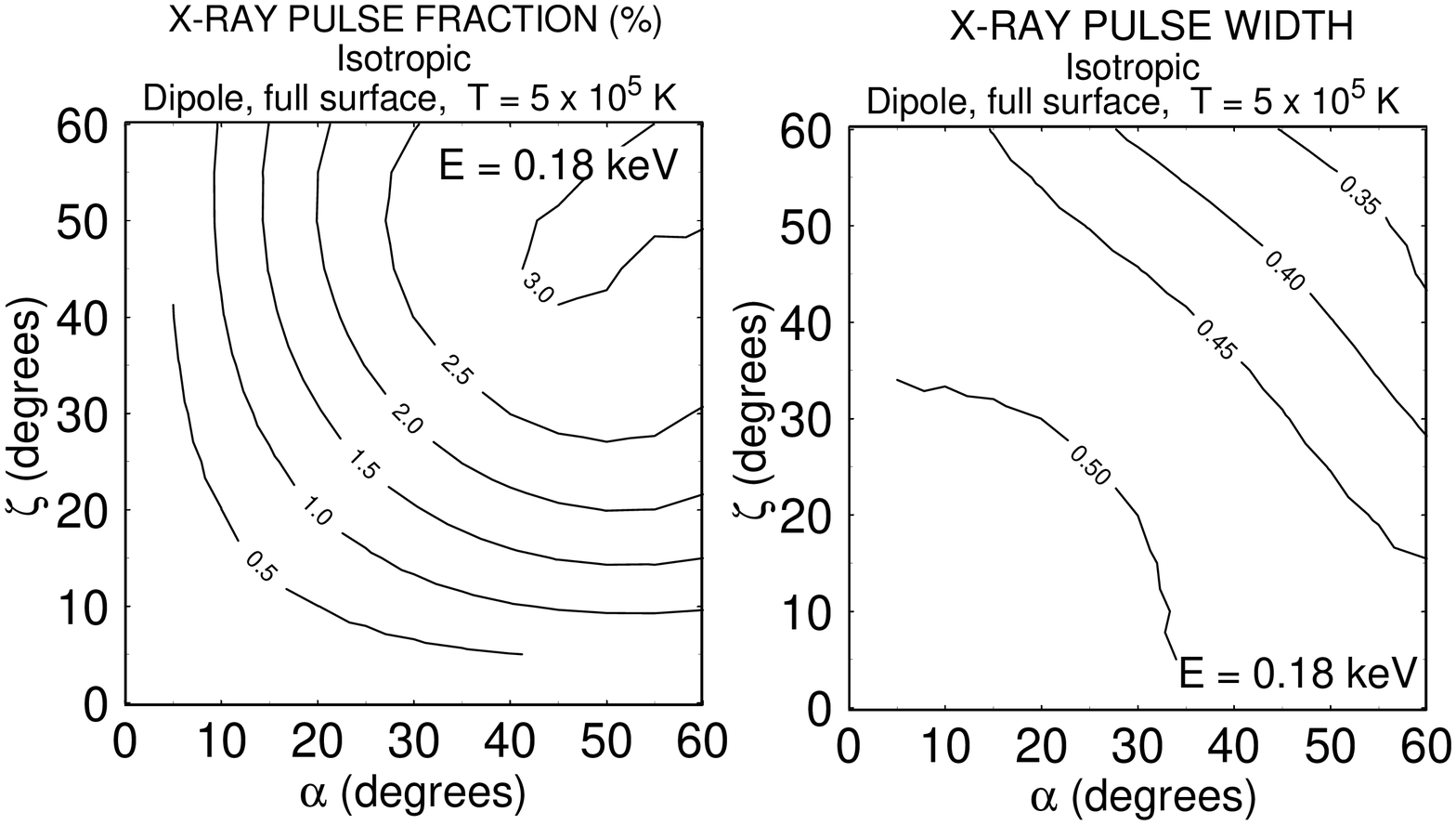}{0}{
Contours of constant X-ray pulsed fraction (\%) (left panel) 
and pulse width (right panel) for the case of isotropic emission. 
The calculations are performed for a dipole surface magnetic field 
and for the effective surface temperature at the magnetic pole 
of $5\times 10^5$ K. See \S~4 for details. 
\label{fig:4}} 

\figureout{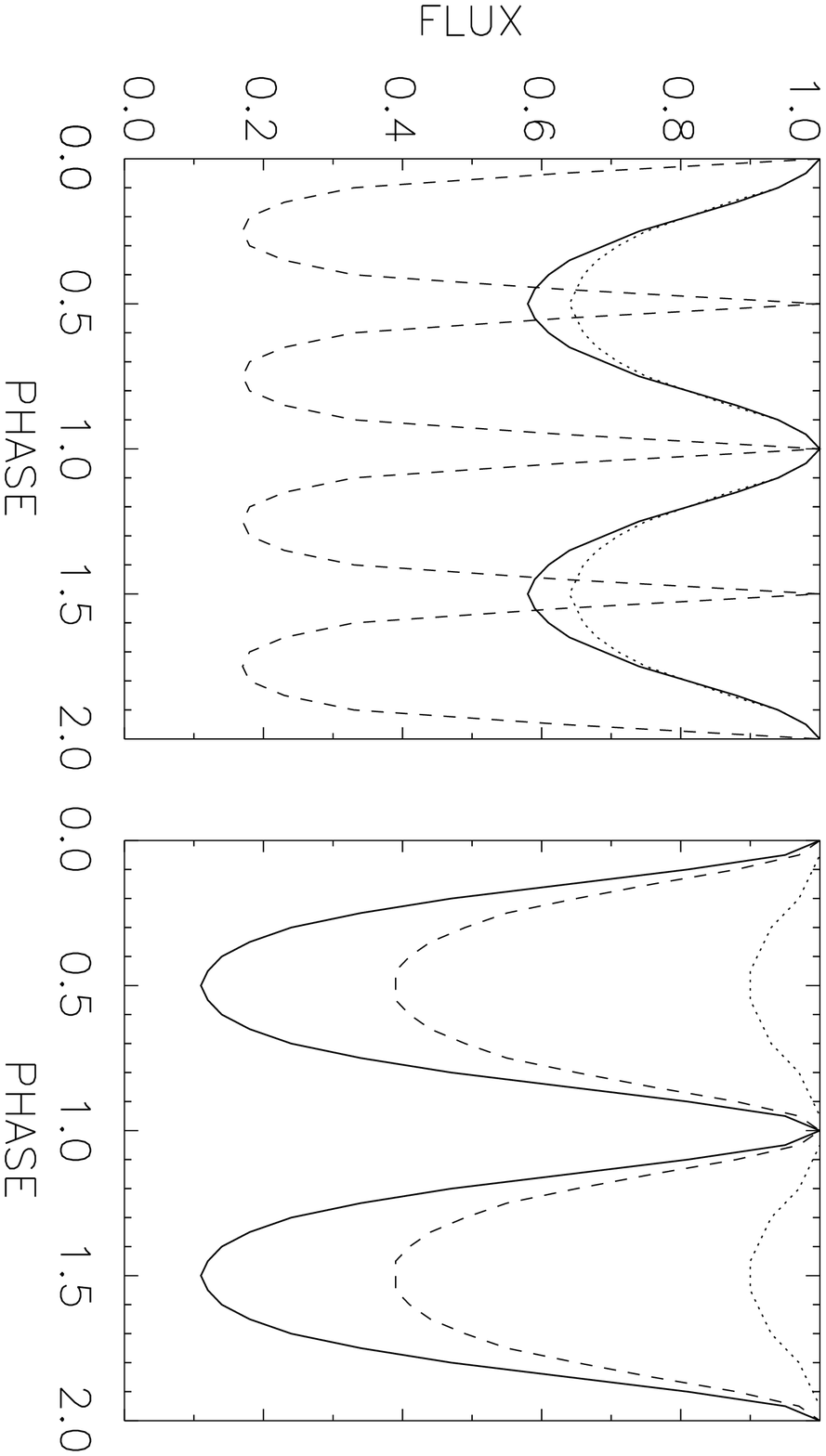}{90}{ 
Left panel: X-ray light curves calculated for the emission 
profile  shown in Figure 3 by a solid line and for different obliquity
and observer angles 
(solid line: $\alpha = \zeta = 10^o$; dotted line: $\alpha =15^o$, 
$\zeta = 60^o$; and dashed line: $\alpha =\zeta = 90^o$). Right panel: 
X-ray light curves calculated for the different emission profiles 
represented by the pencil (solid line), fan (dotted
line), and pencil+fan (dashed line) components of the emission pattern
shown in Figure 3 by a solid line and for $\alpha = \zeta = 15^o$. 
Other details are the same as in Figure 4. 
\label{fig:5}}

\figureout{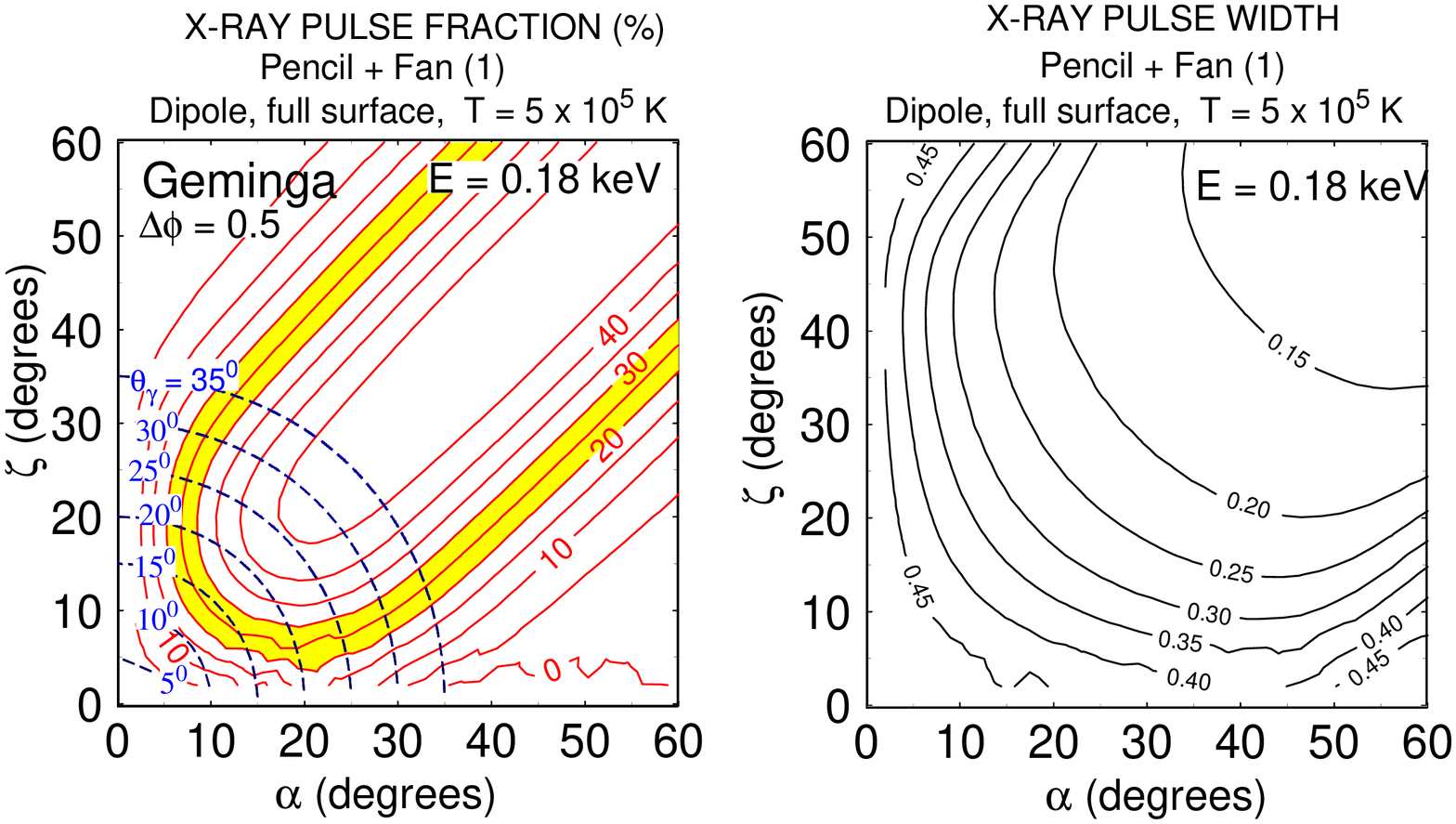}{0}{
Modeling of the soft X-ray and $\gamma $-ray emission for  
Geminga. 
Left panel: contours of constant X-ray pulsed fraction (\%) 
with shaded contours denoting observed pulsed fraction range, 
dashed lines are contours of constant $\gamma $-ray beam 
half-angle (degrees), $\Delta \phi $ is $\gamma $-ray pulse 
phase separation and E is X-ray energy. Right panel: 
contours of constant X-ray pulse width. The 
calculations have been performed for the X-ray emission pattern 
shown in Figure 3 by a solid line. The surface effective temperature 
(at the magnetic pole) $\rm T=5\times 10^5$ K (see \S~4 for details).
\label{fig:6}}

\figureout{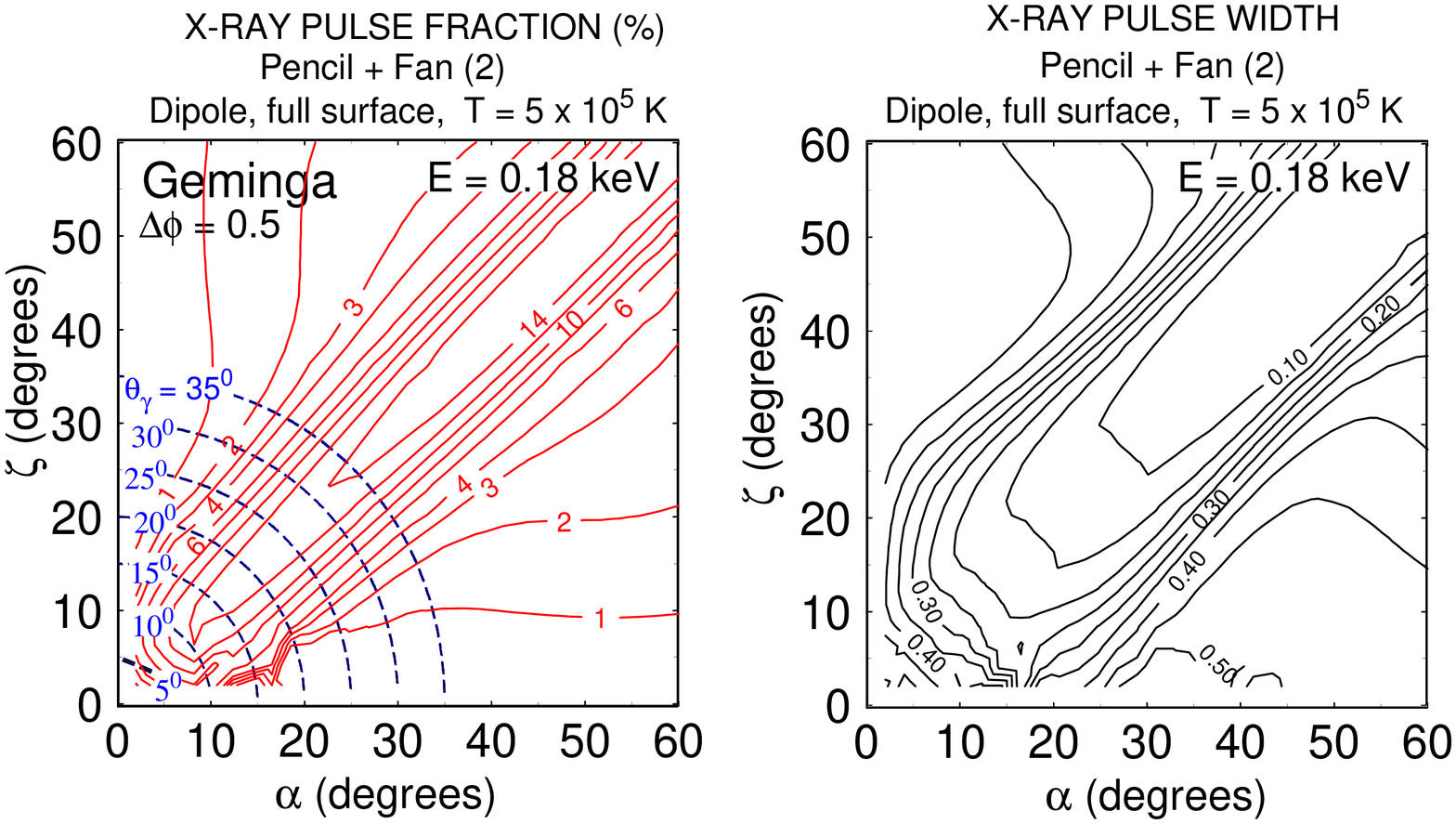}{0}{
Modeling of the soft X-ray and $\gamma $-ray emission for  
Geminga. The calculations have been performed for the X-ray emission 
pattern shown in Figure 3 by dotted line. Other details are the same
as in Figure 6.
\label{fig:7}}

\figureout{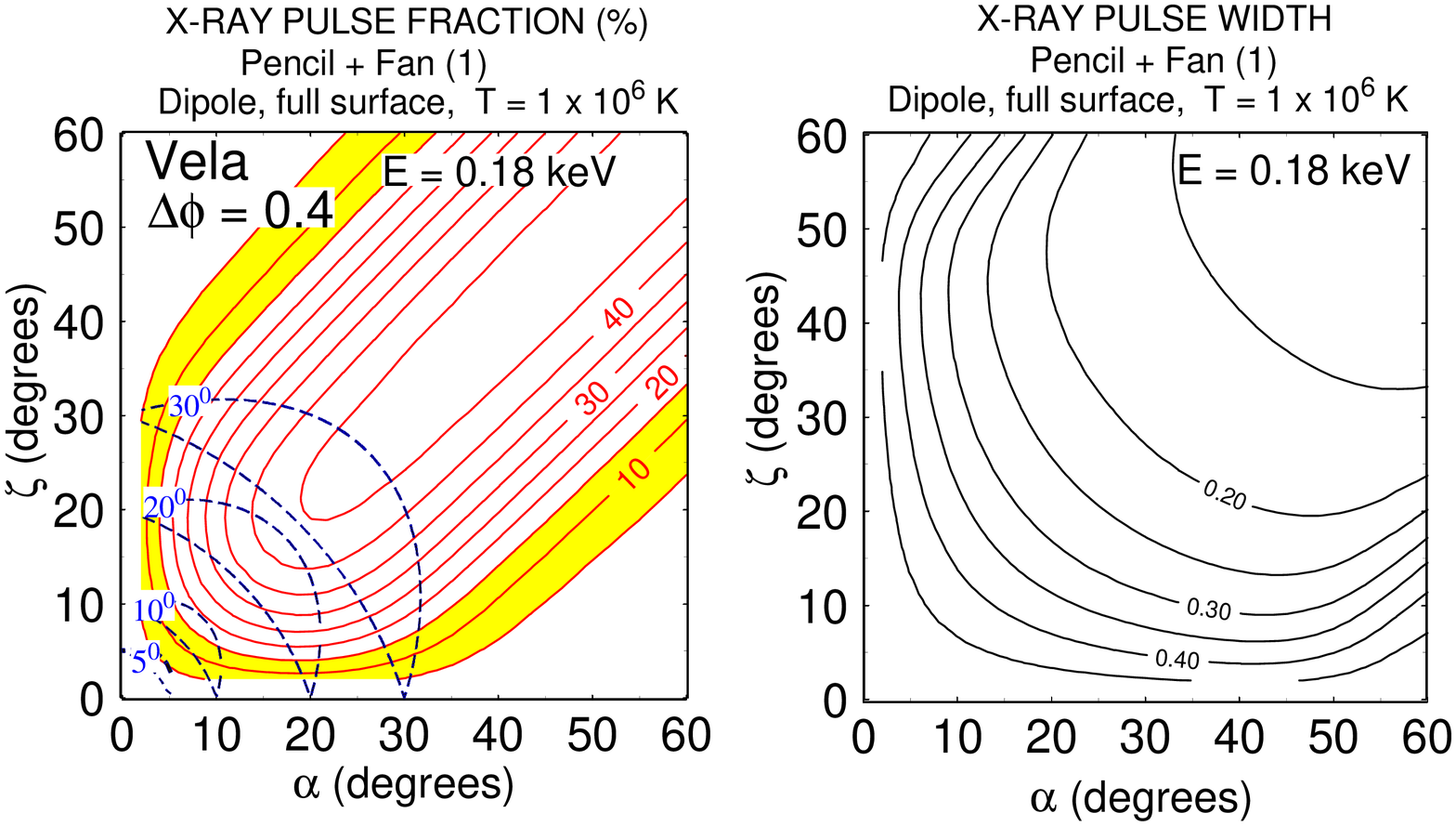}{0}{
Modeling of the soft X-ray and $\gamma $-ray emission for  
Vela (also apply to PSR 0656+14). The surface effective temperature 
at the magnetic pole $\rm T=10^6$ K. Other details are the same as in
Figure 6.    
\label{fig:8}}

\figureout{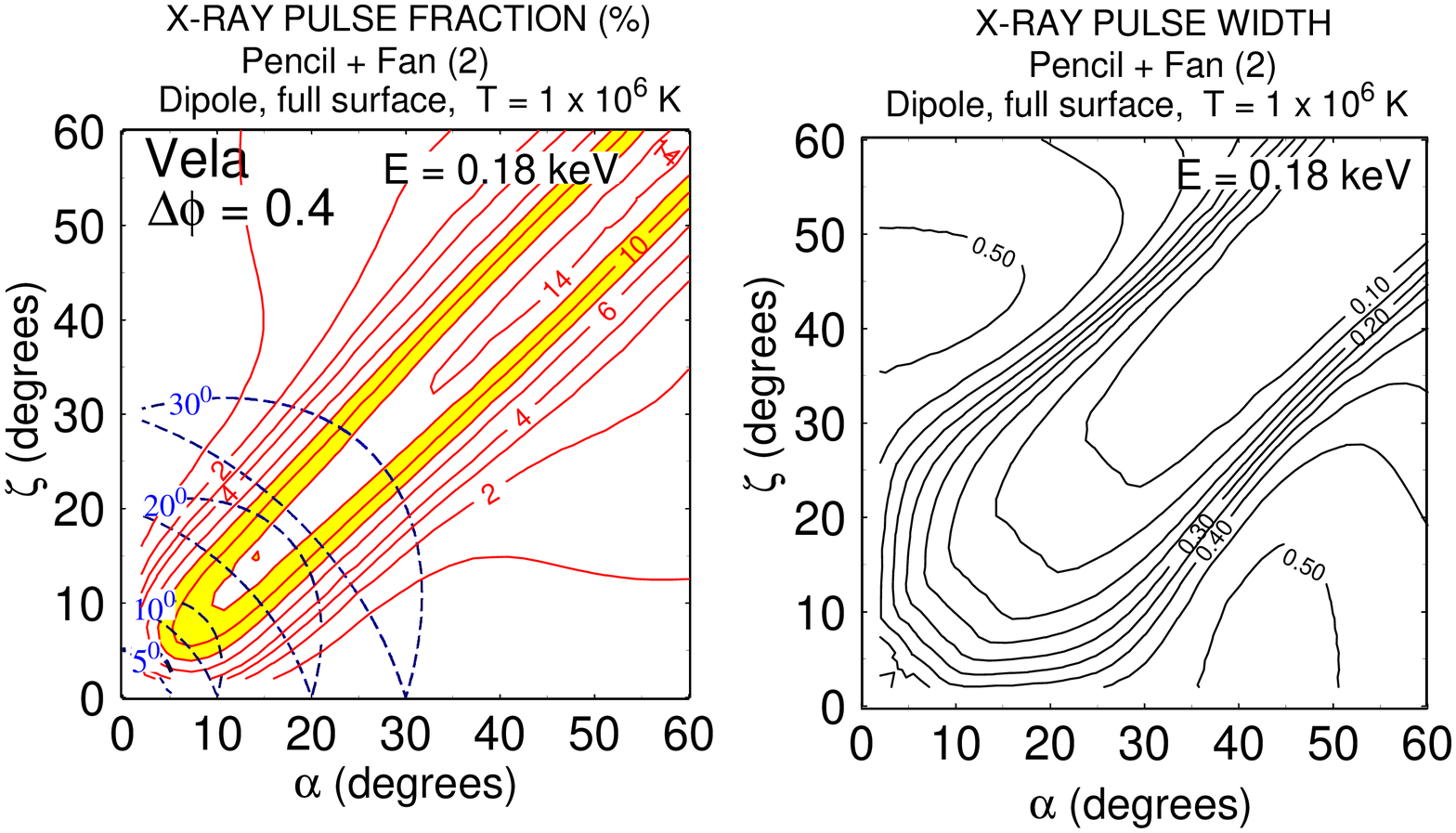}{0}{
Modeling of the soft X-ray and $\gamma $-ray emission for  
Vela (also applies to PSR 0656+14). The surface effective temperature 
(at the magnetic pole) $\rm T = 10^6$ K. Other details are the same
as in Figure 7.
\label{fig:9}} 

\end{document}